\documentclass{ws-procs9x6}

\begin{document}

\title{Status of the LHCb detector\footnote{\uppercase{O}n behalf of the \uppercase{LHC}b collaboration}}

\author{P. Vazquez\footnote{\uppercase{W}ork partially supported by \uppercase{MEYC}
and \uppercase{X}unta de \uppercase{G}alicia}}

\address{Departamento de F\'isica de Part\'iculas, IGFAE \\
Facultade de F\'isica, Campus sur s/n\\
E-15782 Santiago de Compostela, Spain \\
E-mail: fpvazquez@usc.es}

\maketitle

\abstracts{
LHCb is an experiment dedicated to study CP violation and other rare phenomena in B mesons decays with very high precision. It is one of the four experiments that will operate at the 14Tev proton-proton Large Hadron Collider (LHC) at CERN in 2008. Here we briefly describe the current status of the experiment, as well as plans towards a future upgrade.}

\section{Introduction}

The role of LHCb is to search for New Physics (NP) beyond the Standard Model (SM) through precision measurements of CP-violating\cite{gligorov} observables and the study of rare decays\cite{hernando} in the b-quark sector.
LHCb is going to improve the knowledge of all the angles of the so called $\it unitary~triangle$ obtained by B-factories\cite{robbe} (Babar and Belle experiments) and the Tevatron collider, as for example, the $\gamma$ angle which is known with an uncertaninty of $20 \%$. A precise measurement of this angle, by looking at channels such as $B_s\rightarrow D_sK^-$ or $B_{d/s}\rightarrow h^+h^{*-}$ ($h$ and $h^*$ being kaons or pions) could discover NP. In the $B_s$ system, the analysis of $B_s\rightarrow J/\psi\phi$ desintegration will allow the measurement of the $\phi_s$ oscillation phase and a first observation of the rare decay mode $B_s  \rightarrow \mu^+\mu^-$ can be made. The neutral meson oscillation and quark mixing phases are going to be measured to test the CKM matrix and Flavour-Changing Neutral Currents ($b\rightarrow s$) will be seach for.

\begin{figure}[ht]
\centerline{\epsfxsize=4.5in\epsfbox{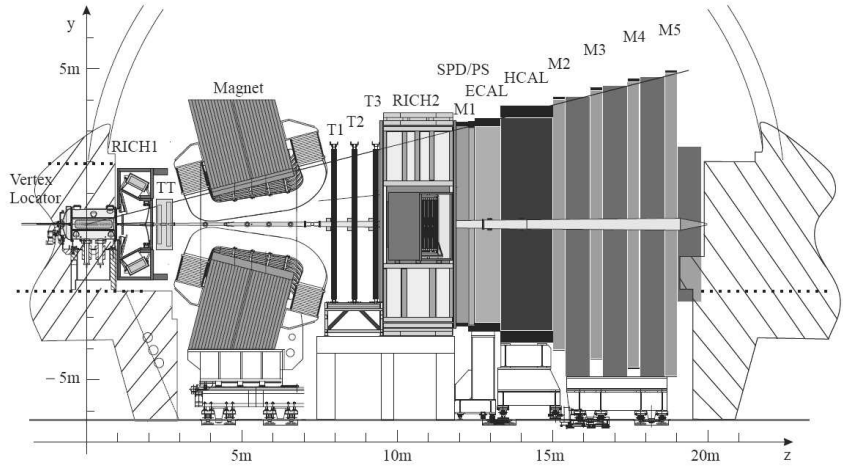}}
\caption{LHCb detector layout. \label{layout}}
\end{figure}

\section{The LHCb detector}

\begin{figure}[ht]
\begin{minipage}[b]{0.48\linewidth}
\epsfxsize=1\linewidth\epsfbox{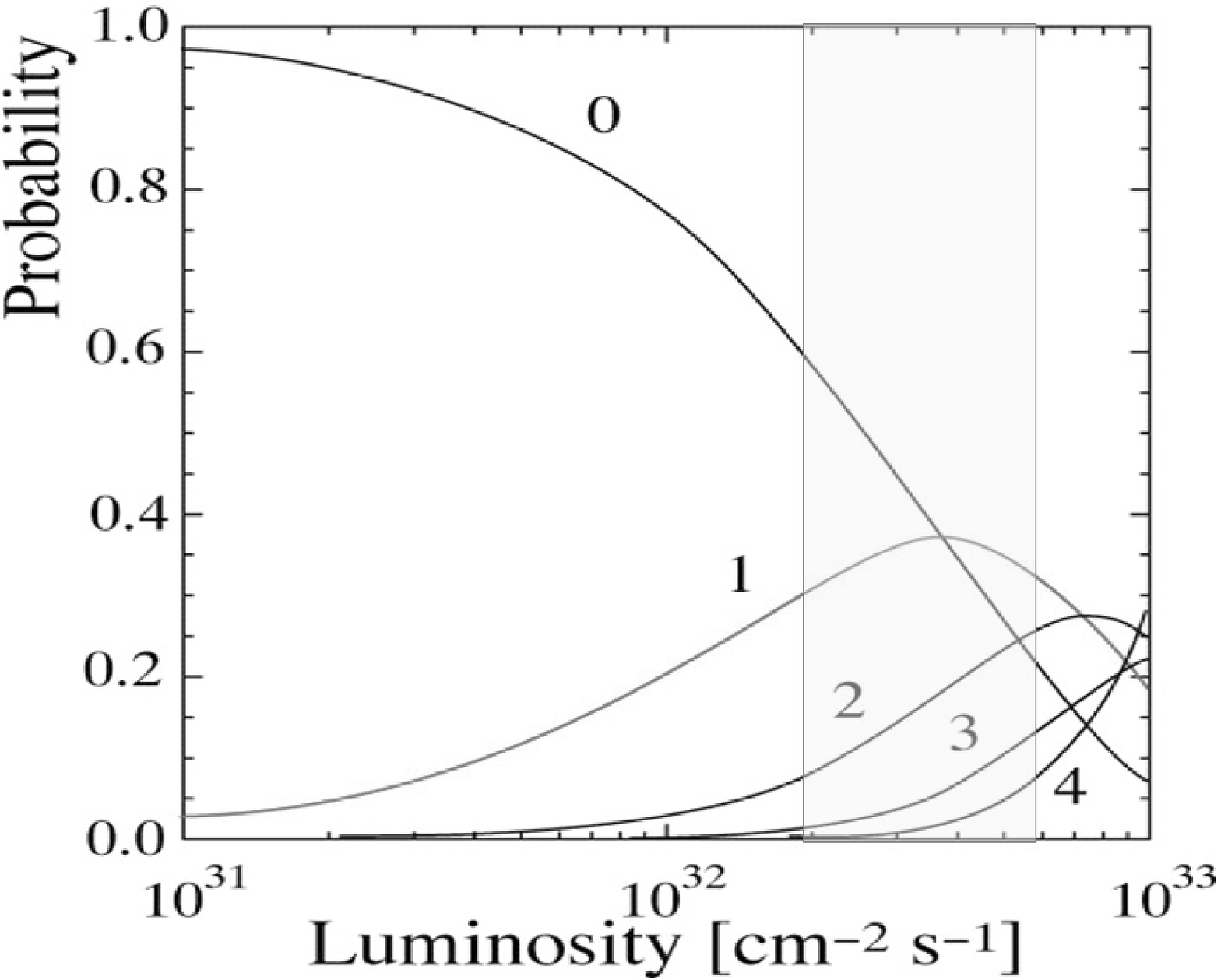}
\caption{pp interactions per bunch crossing versus luminosity at LHCb.
\label{lumi}}
\end{minipage}
\begin{minipage}[b]{0.51\linewidth}
\epsfxsize=1\linewidth\epsfbox{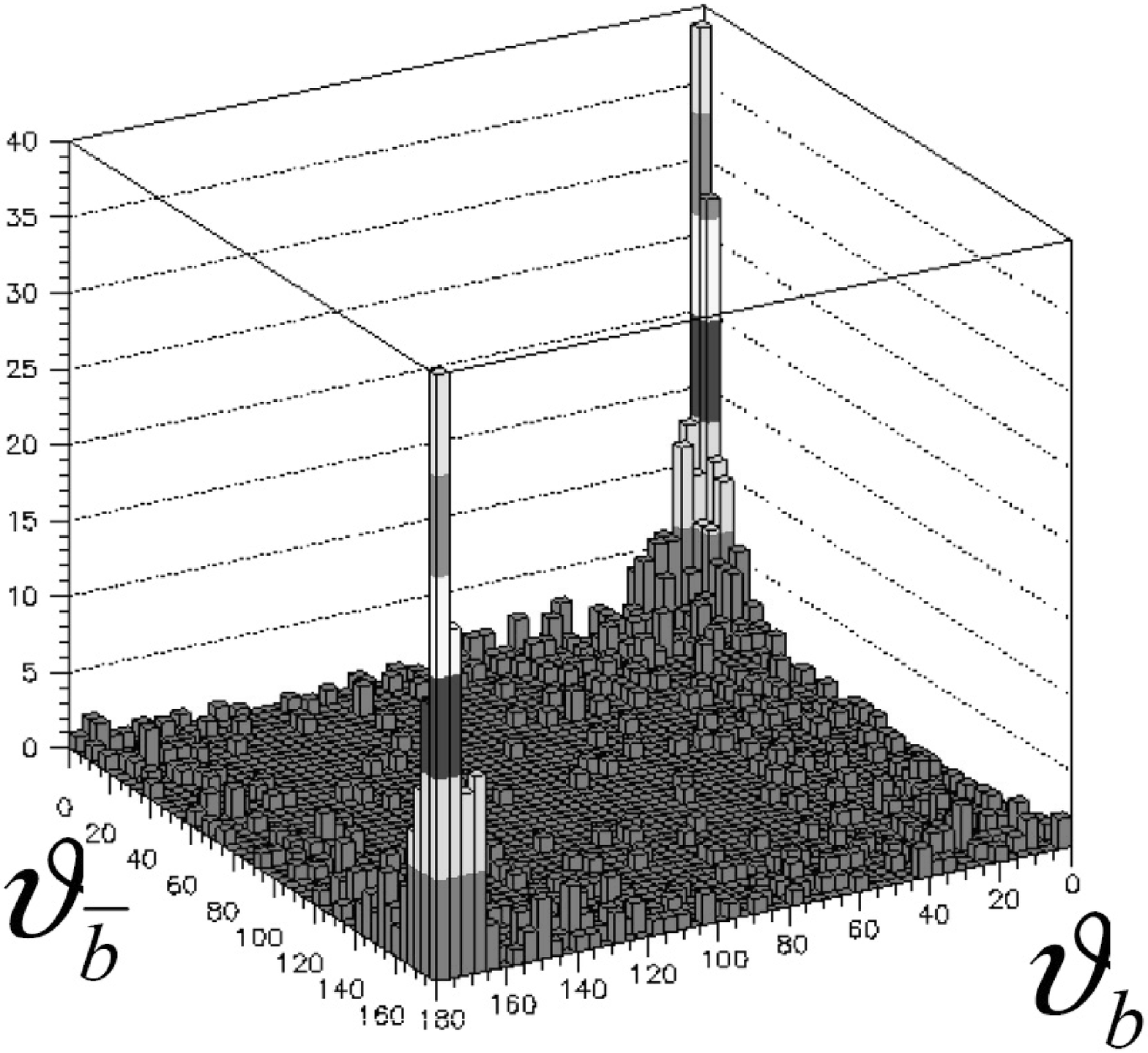}
\caption{$b\bar{b}$ angle with respect to the beam axis at LHCb.
\label{bbangle}}
\end{minipage}
\end{figure}

The LHCb detector\cite{lhcb} is a single-arm spectrometer with the interaction point shifted 11m to the side of the cavern to exploit the forward cone, where the B mesons are intensively momentum resolution as a function of track momentum,produced in the {\it pp} interactions, see Figure \ref{bbangle}. The detector, whose layout can be seen in the Figure \ref{layout}, covers an acceptance of 10-300(15-250) mrad for X(Y) axis. The LHC optics at IP8, where LHCb is placed, was designed to run at a lower luminosity, 2-5x10$^{32}$ cm$^{-2}$s$^{-1}$, as compared to the general purpose experiments at LHC (10$^{34}$ cm$^{-2}$s$^{-1}$),  maximizing the probability of having a single interaction per bunch crossing, see Figure \ref{lumi}, simplifying the event reconstruction, improving the primary to secondary vertex separation and reducing the radiation damage. Even with this lower luminosity it is foreseen that LHCb will produce up to $10^{12}$ $b\bar{b}$ pairs per year ($10^7$ s).

The study of B physics at LHCb is determined by the relatively long lifetime and decay path of the B hadrons, their high mass which produces desintegration products with high $p_t$ and the fact that both b- and $\bar{b}$-quarks produced in a pp interaction fly in the same direction. Therefore in order to tag the corresponding B messons we have to search for high $p_t$ particles displaced from the primary vertex.

\begin{figure}[ht]
\begin{minipage}[b]{0.47\linewidth}
\epsfxsize=1\linewidth\epsfbox{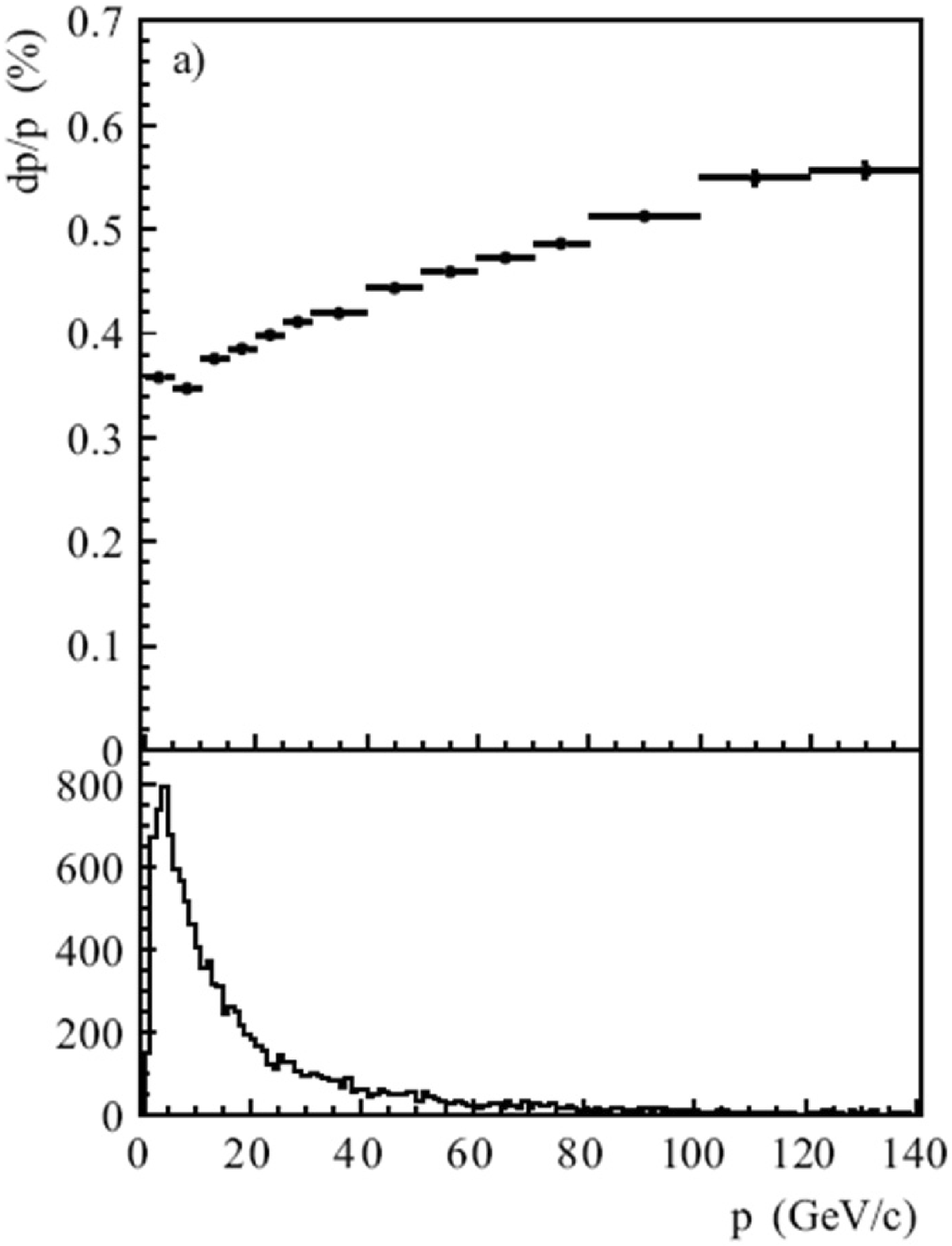}
\caption{Momentum resolution (high) and spectra (low) as a function of $p$.
\label{dpp}}
\end{minipage}
\begin{minipage}[b]{0.52\linewidth}
\epsfxsize=1\linewidth\epsfbox{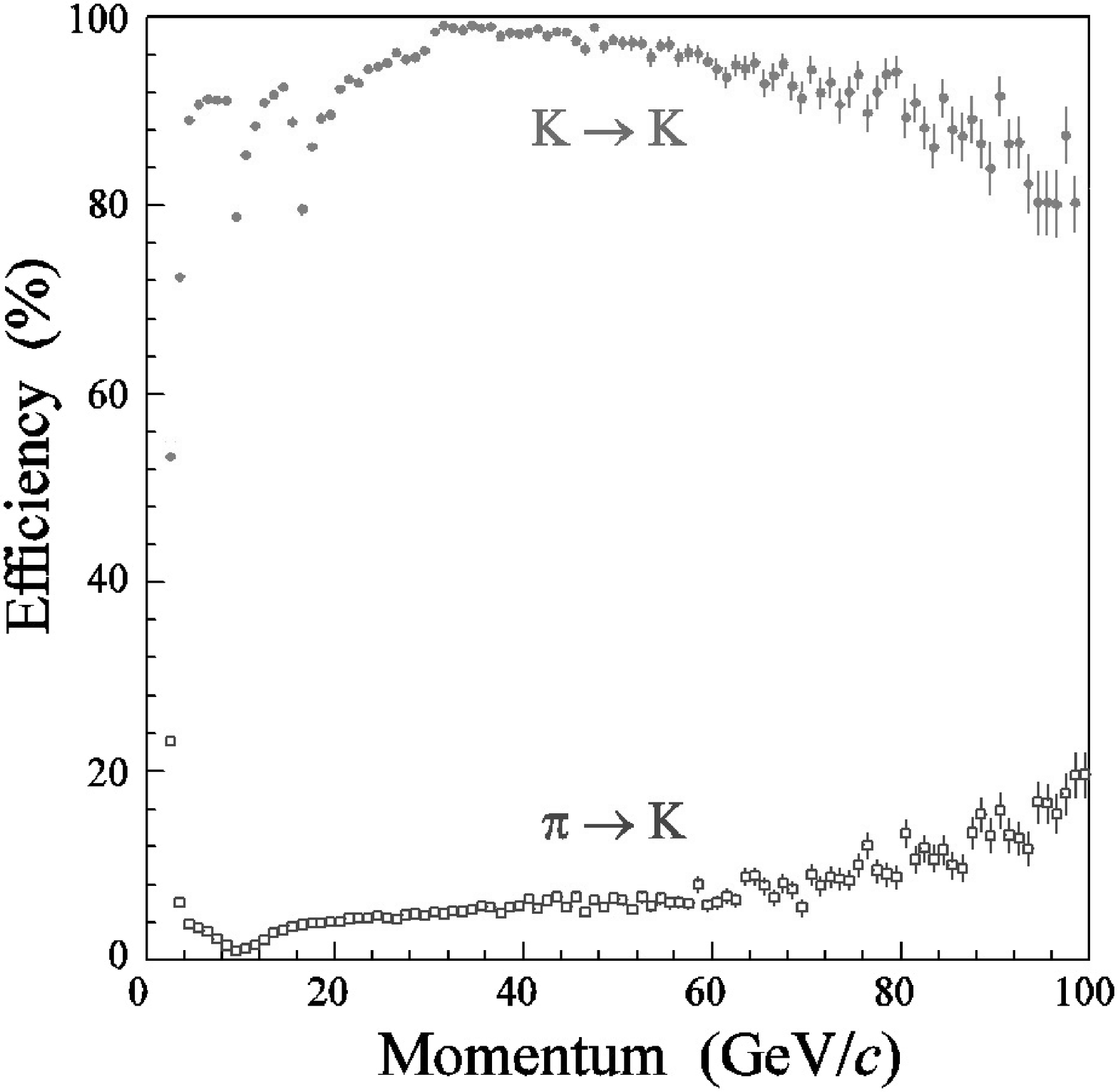}
\caption{Kaon identification and pion missidentification for the LHCb RICH system.
\label{pk}}
\end{minipage}
\end{figure}

The LHCb detector consists of a vertex locator (VELO)\cite{velo}, a 4Tm warm dipole magnet, a tracking system (TT, IT\cite{st}, OT\cite{ot}), two ring imaging cherenkov detectors (RICH1, RICH2)\cite{rich}, a calorimeter system\cite{calo} and a muon system\cite{muon}. The VELO is comprised of 21 stations of two semi-circular silicon sensors measuring the radial and the azimuthal coordinates. The sensors can be placed at 8mm to the beam line during stable conditions or can be retracted during beam injection. The TT consists of four layers of 200 $\mu$m pitch silicon sensors placed upstream of the magnet. The three tracking stations downstream of the magnet are equipped with silicon sensors in the inner part (IT), a cross shaped area around the beam pipe that covers 2\% of the acceptance and gets 20\% of tracks, and with drift time tubes in the outer part (OT). The tracking system is able to measure the impact parameter within an error of 30 $\mu$m and the momentum with $0.4\%$, see Figure \ref{dpp}. RICH1 detector provides particle identification upstream of the magent from 1-60 GeV using aerogel and $C_4F_{10}$ as radiators. RICH2 identifies particles in the momentum range 15-100 GeV downstream of the magnet using $CF_4$. The kaon identification efficiency is 88\% and the pion missindentification is 3\%, see Figure \ref{pk}. The calorimeter system: a preshower detector, electronic and hadronic calorimeters that select high $E_t$ hadron, electron and photon candidates for the L0-trigger. The five muon stations provide fast information for the high $p_t$ for the L0-trigger and muon identification for the High Level trigger. The installation of all detectors is basically complete and most of them are being commissioned using cosmics where possible, otherwhise LED or test pulses.

The LHCb trigger\cite{trigger} is divided in two levels, the Level-0 (L0) hardware implemented and the High Level Trigger (HLT) software implemented. The L0 reduces the rate of bunch crossings with interactions to below a rate of 1.1 MHz, the maximum rate at which the frontend electronics can be readout. It reconstructs the highest $E_t$ hadron, electron and photon, and the two highest $p_t$ muons. After L0, all detectors are readout, and full event building is performed on the CPU nodes of the event filter farm. The HLT consists of a C++ aplication which is running on every CPU and limits the events being written to storage to a rate of $\sim$ 2 kHz. The typical efficiency of the whole trigger chain for hadronic, radiative and leptonic B-decays is 25-30\%, 30-40\% and 60-70\% respectively.

\section{The LHCb upgrade}

It is expected that the LHC will deliver the first pp colisions in summer 2008. The aim of LHCb is to look for NP signatures compatible with luminosities around 0.5 fb$^{-1}$. Within the next four to five years, it will accumulate $\sim$ 10 fb$^{-1}$, enough to make a first observation of rare decays $B_s \rightarrow \mu^+\mu^-$ down to the SM prediction and improve CKM angle $\gamma$ by a factor 5. To continue at the nominal luminosity  will not be very profitable as statistical precision increases very slowly. Therefore LHCb is in the process of studying upgrade\cite{upgrade} possibilities which would allow to collect up to 100 fb$^{-1}$ and would be implemented around 2014. To increase the luminosity at IP8 by a factor ten, up to 2x10$^{33}$ cm$^{-2}$s$^{-1}$, is not a concern so the upgrade of LHCb is purely a question of the detector being able to profit from a higher peak luminosity. The main worries are that all the sub-detectors need to be readout at a 40 MHz, requiring that the frontend electronics of all them, with the exception of the muon, need to be replaced; silicon detectors need to be replaced in the area close to the beam pipe, with a possible enlargement of the IT acceptance, in order to have more granularity and keep being radiation tolerant; all trigger decisions will be done in the computer farm.

\section{Summary}

After 13 years of work the LHCb detector will be ready for searching for New Physics when the LHC turns on in summer 2008. A new era for the flavour physics is open.


\begin{thebibliography}{0}

\bibitem{gligorov}V. Gligorov,
{\it Prospects for CP Violation Studies at LHCb}, these proceedings.

\bibitem{hernando}J.A. Hernando,
{\it Prospects of rare B mesons decays studies with the LHCb experiment}, these proceedings.

\bibitem{robbe}P. Robbe,
{\it Nucl. Phys. B (Proc. Suppl.)}, {\bf 170}, 46 (2007).
 
\bibitem{lhcb}LHCb collaboration, [R. Antunes Nobrega et al.],
{\it LHCb Reoptimized Detector Design and Performance}, CERN-LHCC-2003-030.

\bibitem{velo}LHCb collaboration, [P. R. Barbosa-Marinho et al.],
{\it LHCb VELO (VErtex LOcator) : Technical Design Report}, CERN-LHCC-2001-011.

\bibitem{st}LHCb collaboration, [P. R. Barbosa-Marinho et al.],
{\it LHCb inner tracker : Technical Design Report}, CERN-LHCC-2002-029.

\bibitem{ot}LHCb collaboration, [P. R. Barbosa-Marinho et al.],
{\it LHCb outer tracker : Technical Design Report}, CERN-LHCC-2001-024.

\bibitem{rich}LHCb collaboration, [S. Amato et al.],
{\it LHCb RICH : Technical Design Report}, CERN-LHCC-2000-037.

\bibitem{calo}LHCb collaboration, [S. Amato et al.],
{\it LHCb calorimeters : Technical Design Report}, CERN-LHCC-2000-036.

\bibitem{muon}LHCb collaboration, [P. R. Barbosa-Marinho et al.],
{\it LHCb muon system : Technical Design Report}, CERN-LHCC-2001-010.

\bibitem{trigger}LHCb collaboration, [R. Antunes Nobrega et al.],
{\it LHCb Trigger System : Technical Design Report}, CERN-LHCC-2003-031.

\bibitem{upgrade}LHCb collaboration,
{\it Expression of Interest for an LHCb Upgrade}, CERN-LHCb-2008-019.

\end{thebibliography}
\end{document}